# Multi-Shot Distributed Transaction Commit

## Gregory Chockler
Royal Holloway, University of London, UK

## Alexey Gotsman[1]
IMDEA Software Institute, Madrid, Spain

**Abstract**

Atomic Commit Problem (ACP) is a single-shot agreement problem similar to consensus, meant to model the properties of transaction commit protocols in fault-prone distributed systems. We argue that ACP is too restrictive to capture the complexities of modern transactional data stores, where commit protocols are integrated with concurrency control, and their executions for different transactions are interdependent. As an alternative, we introduce Transaction Certification Service (TCS), a new formal problem that captures safety guarantees of multi-shot transaction commit protocols with integrated concurrency control. TCS is parameterized by a certification function that can be instantiated to support common isolation levels, such as serializability and snapshot isolation. We then derive a provably correct crash-resilient protocol for implementing TCS through successive refinement. Our protocol achieves a better time complexity than mainstream approaches that layer two-phase commit on top of Paxos-style replication.

# 1 Introduction

Modern data stores are often required to manage massive amounts of data while providing stringent transactional guarantees to their users. They achieve scalability by partitioning data into independently managed *shards* (aka *partitions*) and fault-tolerance by replicating each shard across a set of servers [8, 13, 33, 41]. Implementing such systems requires sophisticated protocols to ensure that distributed transactions satisfy a conjunction of desirable properties commonly known as ACID: Atomicity, Consistency, Isolation and Durability.

Traditionally, distributed computing literature abstracts ways of achieving these properties into separate problems: in particular, atomic commit problem (ACP) for Atomicity and concurrency control (CC) for Isolation. ACP is formalised as a *one-shot* agreement problem in which multiple shards involved in a transaction need to reach a decision on its final outcome: COMMIT if all shards voted to commit the transaction, and ABORT otherwise [12]. Concurrency control is responsible for determining whether a shard should vote to commit or abort a transaction based on the locally observed conflicts with other active transactions. Although both ACP and CC must be solved in any realistic transaction processing system, they are traditionally viewed as disjoint in the existing literature. In particular, solutions for ACP treat the votes as the inputs of the problem, and leave the interaction with CC, which is responsible for generating the votes, outside the problem scope [2, 15, 22, 37].

---

[1] Alexey Gotsman was supported by an ERC Starting Grant RACCOON.

This separation, however, is too simplistic to capture the complexities of many practical implementations in which commit protocols and concurrency control are tightly integrated, and as a result, may influence each other in subtle ways. For example, consider the classical *two-phase commit (2PC)* protocol [14] for solving ACP among reliable processes. A transaction processing system typically executes a 2PC instance for each transaction [30, 31, 33, 38]. When a processes $p_i$ managing a shard $s$ receives a transaction $t$, it performs a local concurrency-control check and accordingly votes to commit or abort $t$. The votes on $t$ by different processes are aggregated, and the final decision is then distributed to all processes. If $p_i$ votes to commit $t$, as long as it does not know the final decision on $t$, it will have to conservatively presume $t$ as committed. This may cause $p_i$ to vote ABORT in another 2PC instance for a transaction $t'$ conflicting with $t$, even if in the end $t$ is aborted. In this case, the outcome of one 2PC instance (for $t'$) depends on the internals of the execution of another instance (for $t$) and the concurrency-control policy used.

At present, the lack of a formal framework capturing such intricate aspects of real implementations makes them difficult to understand and prove correct. In this paper, we take the first step towards bridging this gap. We introduce *Transaction Certification Service (TCS, §2)*, a new formal problem capturing the safety guarantees of a multi-shot transaction commit protocol with integrated concurrency control. The TCS exposes a simple interface allowing clients to submit transactions for *certification* via a `certify` request, which returns COMMIT or ABORT. A TCS is meant to be used in the context of transactional processing systems with optimistic concurrency control, where transactions are first executed optimistically, and the results (e.g., read and write sets) are submitted for certification to the TCS. In contrast to ACP, TCS does not impose any restrictions on the number of repeated `certify` invocations or their concurrency. It therefore lends itself naturally to formalising the interactions between transaction commit and concurrency control. To this end, TCS is parameterised by a *certification function*, which encapsulates the concurrency-control policy for the desired isolation level, such as serializability and snapshot isolation [1]. The correctness of TCS is then formulated by requiring that its certification decisions be consistent with the certification function.

We leverage TCS to develop a formal framework for constructing provably correct multi-shot transaction commit protocols with customisable isolation levels. The core ingredient of our framework is a new *multi-shot two-phase commit protocol* (§3). It formalises how the classical 2PC interacts with concurrency control in many practical transaction processing systems [30, 31, 33, 38] in a way that is parametric in the isolation level provided. The protocol also serves as a *template* for deriving more complex TCS implementations. We prove that the multi-shot 2PC protocol correctly implements a TCS with a given certification function, provided the concurrency-control policies used by each shard match this function.

We next propose a *crash fault-tolerant* TCS implementation and establish its correctness by proving that it simulates multi-shot 2PC (§4). A common approach to making 2PC fault-tolerant is to get every shard to simulate a reliable 2PC process using a replication protocol, such as Paxos [8, 13, 15, 17, 41]. Similarly to recent work [25, 40], our implementation optimises the time complexity of this scheme by weaving 2PC and Paxos together. In contrast to previous work, our protocol is both generic in the isolation level and rigorously proven correct. It can therefore serve as a reference solution for future distributed transaction commit implementations. Moreover, a variant of our protocol has a time complexity matching the lower bounds for consensus [6, 23] and non-blocking atomic commit [12].

The main idea for achieving such a low time complexity is to eliminate the Paxos consensus required in the vanilla fault-tolerant 2PC to persist the final decision on a transaction at a shard. Instead, the decision is propagated to the relevant shard replicas asynchronously. This means that different shard replicas may receive the final decision on a transaction at different times, and thus their states may be inconsistent. To deal with this, in our protocol the votes are computed locally by a single shard *leader* based on the information available to it; other processes merely store the votes. Similarly to [21, 28], such a *passive replication* approach requires a careful design of recovery from leader failures. Another reduction in time complexity comes from the fact that our protocol avoids consistently replicating the 2PC coordinator: we allow any process to take over as a coordinator by accessing the current state of the computation at shards. The protocol ensures that all coordinators will reach the same decision on a transaction.

## 2 Transaction Certification Service

**Interface.** A *Transaction Certification Service (TCS)* accepts *transactions* from $\mathcal{T}$ and produces *decisions* from $\mathcal{D} = \{\text{ABORT}, \text{COMMIT}\}$. Clients interact with the TCS using two types of *actions*: certification requests of the form $\texttt{certify}(t)$, where $t \in \mathcal{T}$, and responses of the form $\texttt{decide}(t, d)$, where $d \in \mathcal{D}$.

In this paper we focus on transactional processing systems using optimistic concurrency control. Hence, we assume that a transaction submitted to the TCS includes all the information produced by its optimistic execution. As an example, consider a transactional system managing *objects* in the set Obj with values in the set Val, where transactions can execute reads and writes on the objects. The objects are associated with a totally ordered set Ver of *versions* with a distinguished minimum version $v_0$. Then each transaction $t$ submitted to the TCS may be associated with the following data:
- *Read set* $R(t) \subseteq 2^{\text{Obj} \times \text{Ver}}$: the set of objects with their versions that $t$ read, which contains at most one version per object.
- *Write set* of $W(t) \subseteq 2^{\text{Obj} \times \text{Val}}$: the set of objects with their values that $t$ wrote, which contains at most one value per object. We require that any object written has also been read: $\forall (x, \_) \in W(t). (x, \_) \in R(t)$.
- *Commit version* $V_c(t) \in \text{Ver}$: the version to be assigned to the writes of $t$. We require that this version be higher than any of the versions read: $\forall (\_, v) \in R(t). V_c(t) > v$.

**Certification functions.** A TCS is specified using a *certification function* $f : 2^{\mathcal{T}} \times \mathcal{T} \to \mathcal{D}$, which encapsulates the concurrency-control policy for the desired isolation level. The result $f(T, t)$ is the decision for the transaction $t$ given the set of the previously committed transactions $T$. We require $f$ to be *distributive* in the following sense:

$$\forall T_1, T_2, t. f(T_1 \cup T_2, t) = f(T_1, t) \sqcap f(T_2, t), \tag{1}$$

where the $\sqcap$ operator is defined as follows: COMMIT $\sqcap$ COMMIT = COMMIT and $d \sqcap$ ABORT = ABORT for any $d$. This requirement is justified by the fact that common definitions of $f(T, t)$ check $t$ for conflicts against each transaction in $T$ separately.

For example, given the above domain of transactions, the following certification function encapsulates the classical concurrency-control policy for serializability [39]: $f(T, t) = $ COMMIT iff none of the versions read by $t$ have been overwritten by a transaction in $T$, i.e.,

$$\forall x, v. (x, v) \in R(t) \implies (\forall t' \in T. (x, \_) \in W(t') \implies V_c(t') \leq v). \tag{2}$$

A certification function for snapshot isolation (SI) [1] is similar, but restricts the certification check to the objects the transaction $t$ writes: $f(T, t) = $ COMMIT iff

$$\forall x, v. (x, v) \in R(t) \land (x, \_) \in W(t) \implies (\forall t' \in T. (x, \_) \in W(t') \implies V_c(t') \leq v). \tag{3}$$

It is easy to check that the certification functions (2) and (3) are distributive.

**Histories.** We represent TCS executions using *histories*—sequences of `certify` and `decide` actions such that every transaction appears at most once as a parameter to `certify`, and each `decide` action is a response to exactly one preceding `certify` action. For a history $h$ we let $\mathsf{act}(h)$ be the set of actions in $h$. For actions $a, a' \in \mathsf{act}(h)$, we write $a \prec_h a'$ when $a$ occurs before $a'$ in $h$. A history $h$ is *complete* if every `certify` action in it has a matching `decide` action. A complete history is *sequential* if it consists of pairs of `certify` and matching `decide` actions. A transaction $t$ *commits* in a history $h$ if $h$ contains $\mathsf{decide}(t, \textsc{commit})$. We denote by $\mathsf{committed}(h)$ the projection of $h$ to actions corresponding to the transactions that are committed in $h$. For a complete history $h$, a *linearization* $\ell$ of $h$ [20] is a sequential history such that: *(i)* $h$ and $\ell$ contain the same actions; and *(ii)*

$$\forall t, t'. \, \mathsf{decide}(t, \_) \prec_h \mathsf{certify}(t') \implies \mathsf{decide}(t, \_) \prec_\ell \mathsf{certify}(t').$$

**TCS correctness.** A complete sequential history $h$ is *legal* with respect to a certification function $f$, if its certification decisions are computed according to $f$:

$$\forall a = \mathsf{decide}(t, d) \in \mathsf{act}(h). \, d = f(\{t' \mid \mathsf{decide}(t', \textsc{commit}) \prec_h a\}, t).$$

A history $h$ is *correct* with respect to $f$ if $h \mid \mathsf{committed}(h)$ has a legal linearization. A TCS implementation is *correct* with respect to $f$ if so are all its histories.

A correct TCS can be readily used in a transaction processing system. For example, consider the domain of transactions defined earlier. A typical system based on optimistic concurrency control will ensure that transactions submitted for certification read versions that already exist in the database. Formally, it will produce only histories $h$ such that, for a transaction $t$ submitted for certification in $h$, if $(x, v) \in R(t)$, then there exists a $t'$ such that $(x, v) \in W(t')$, and $h$ contains $\mathsf{decide}(t', \textsc{commit})$ before $\mathsf{certify}(t)$. It is easy to check that, if such a history $h$ is correct with respect to the certification function (2), then it is also serializable. Hence, TCS correct with respect to certification function (2) can indeed be used to implement serializability.

## 3  Multi-Shot 2PC and Shard-Local Certification Functions

We now present a multi-shot version of the classical *two-phase commit (2PC)* protocol [14], parametric in the concurrency-control policy used by each shard. We then prove that the protocol implements a correct transaction certification service parameterised by a given certification function, provided per-shard concurrency control matches this function. Like 2PC, our protocol assumes reliable processes. In the next section, we establish the correctness of a protocol that allows crashes by proving that it simulates the behaviour of multi-shot 2PC.

**System model.** We consider an asynchronous message-passing system consisting of a set of processes $\mathcal{P}$. In this section we assume that processes are reliable and are connected by reliable FIFO channels. We assume a function $\mathsf{client} : \mathcal{T} \to \mathcal{P}$ determining the client process that issued a given transaction. The data managed by the system are partitioned into *shards* from a set $\mathcal{S}$. A function $\mathsf{shards} : \mathcal{T} \to 2^\mathcal{S}$ determines the shards that need to certify a given transaction, which are usually the shards storing the data the transaction accesses. Each shard $s \in \mathcal{S}$ is managed by a process $\mathsf{proc}(s) \in \mathcal{P}$. For simplicity, we assume that different processes manage different shards.

```
1  next ← −1 ∈ ℤ;
2  txn[ ] ∈ ℕ → 𝒯;
3  vote[ ] ∈ ℕ → {COMMIT, ABORT};
4  dec[ ] ∈ ℕ → {COMMIT, ABORT};
5  phase[ ] ← (λk. START) ∈ ℕ → {START, PREPARED, DECIDED};

6  function certify(t)
7  |   send PREPARE(t) to proc(shards(t));

8  when received PREPARE(t)
9  |   next ← next + 1;
10 |   txn[next] ← t;
11 |   vote[next] ← f_{s_0}({txn[k] | k < next ∧ phase[k] = DECIDED ∧ dec[k] = COMMIT}, t) ⊓
   |                g_{s_0}({txn[k] | k < next ∧ phase[k] = PREPARED ∧ vote[k] = COMMIT}, t);
12 |   phase[next] ← PREPARED;
13 |   send PREPARE_ACK(s_0, next, t, vote[next]) to coord(t);

14 when received PREPARE_ACK(s, pos_s, t, d_s) for every s ∈ shards(t)
15 |   send DECISION(t, ⊓_{s∈shards(t)} d_s) to client(t);
16 |   forall s ∈ shards(t) do
17 |   |   send DECISION(pos_s, ⊓_{s∈shards(t)} d_s) to proc(s)

18 when received DECISION(k, d)
19 |   dec[k] ← d;
20 |   phase[k] ← DECIDED;

21 non-deterministically for some k ∈ ℕ
22 |   pre: phase[k] = DECIDED;
23 |   phase[k] ← PREPARED;

24 non-deterministically for some k ∈ ℕ
25 |   pre: phase[k] ≠ START;
26 |   send PREPARE_ACK(s_0, k, txn[t], vote[k]) to coord(t);
```

**Figure 1** Multi-shot 2PC protocol at a process $p_i$ managing a shard $s_0$.

**Protocol: common case.** We give the pseudocode of the protocol in Figure 1 and illustrate its message flow in Figure 2a. Each handler in Figure 1 is executed atomically.

To certify a transaction $t$, a client sends it in a PREPARE message to the relevant shards (line 6)[2]. A process managing a shard arranges all transactions received into a total *certification order*, stored in an array txn; a next variable points to the last filled slot in the array. Upon receiving a transaction $t$ (line 8), the process stores $t$ in the next free slot of txn. The process also computes its *vote*, saying whether to COMMIT or ABORT the transaction, and stores it in an array vote. We explain the vote computation in the following; intuitively, the vote is determined by whether the transaction $t$ conflicts with a previously received transaction. After the process managing a shard $s$ receives $t$, we say that $t$ is *prepared* at $s$. The process keeps track of transaction status in an array phase, whose entries initially store START, and are changed to PREPARED once the transaction is prepared. Having prepared the transaction $t$, the process sends a `PREPARE_ACK` message with its position in the certification order and the vote to a *coordinator* of $t$. This is a process determined using a function $\mathsf{coord} : \mathcal{T} \to \mathcal{P}$ such that $\forall t.\, \mathsf{coord}(t) \in \mathsf{proc}(\mathsf{shards}(t))$.

---

[2] In practice, the client only needs to send the data relevant to the corresponding shard. We omit this optimisation to simplify notation.

The coordinator of a transaction $t$ acts once it receives a `PREPARE_ACK` message for $t$ from each of its shards $s$, which carries the vote $d_s$ by $s$ (line 14). The coordinator computes the final decision on $t$ using the $\sqcap$ operator (§2) and sends it in `DECISION` messages to the client and to all the relevant shards. When a process receives a decision for a transaction (line 18), it stores the decision in a dec array, and advances the transaction's phase to DECIDED.

**Vote computation.** A process managing a shard $s$ computes votes as a conjunction of two *shard-local certification functions* $f_s : 2^{\mathcal{T}} \times \mathcal{T} \to \mathcal{D}$ and $g_s : 2^{\mathcal{T}} \times \mathcal{T} \to \mathcal{D}$. Unlike the certification function of §2, the shard-local functions are meant to check for conflicts only on objects managed by $s$. They take as their first argument the sets of transactions already decided to commit at the shard, and respectively, those that are only prepared to commit (line 11). We require that the above functions be distributive, similarly to (1).

For example, consider the transaction model given in §2 and assume that the set of objects Obj is partitioned among shards: $\mathsf{Obj} = \biguplus_{s \in \mathcal{S}} \mathsf{Obj}_s$. Then the shard-local certification functions for serializability are defined as follows: $f_s(T, t) = \text{COMMIT}$ iff

$$\forall x \in \mathsf{Obj}_s.\, \forall v.\, (x, v) \in R(t) \implies (\forall t' \in T.\, (x, \_) \in W(t') \implies V_c(t') \leq v), \qquad (4)$$

and $g_s(T, t) = \text{COMMIT}$ iff

$$\begin{aligned}\forall x \in \mathsf{Obj}_s.\, \forall v.\, &((x, \_) \in R(t) \implies (\forall t' \in T.\, (x, \_) \notin W(t'))) \land \\ &((x, \_) \in W(t) \implies (\forall t' \in T.\, (x, \_) \notin R(t'))).\end{aligned} \qquad (5)$$

The function $f_s$ certifies a transaction $t$ against previously committed transactions $T$ similarly to the certification function (2), but taking into account only the objects managed by the shard $s$. The function $g_s$ certifies $t$ against transactions $T$ prepared to commit.

The first conjunct of (5) aborts a transaction $t$ if it read an object written by a transaction $t'$ prepared to commit. To motivate this condition, consider the following example. Assume that a shard managing an object $x$ votes to commit a transaction $t'$ that read a version $v_1$ of $x$ and wants to write a version $v_2 > v_1$ of $x$. If the shard now receives another transaction $t$ that read the version $v_1$ of $x$, the shard has to abort $t$: if $t'$ does commit in the end, allowing $t$ to commit would violate serializability, since it would have read stale data. On the other hand, once the shard receives the abort decision on $t'$, it is free to commit $t$.

The second conjunct of (5) aborts a transaction $t$ if it writes to an object read by a transaction $t'$ prepared to commit. To motivate this, consider the following example, adapted from [36]. Assume transactions $t_1$ and $t_2$ both read a version $v_1$ of $x$ at shard $s_1$ and a version $v_2$ of $y$ at shard $s_2$; $t_1$ wants to write a version $v_2' > v_2$ of $y$, and $t_2$ wants to write a version $v_2 > v_1$ of $x$. Assume further that $s_1$ receives $t_1$ first and votes to commit it, and $s_2$ receives $t_2$ first and votes to commit it as well. If $s_1$ now receives $t_2$ and $s_2$ receives $t_1$, the second conjunct of (5) will force them to abort: if the shards let the transactions commit, the resulting execution would not be serializable, since one of the transactions must read the value written by the other.

A simple way of implementing (5) is, when preparing a transaction, to acquire read locks on its read set and write locks on its write set; the transaction is aborted if the locks cannot be acquired. The shard-local certification functions are a more abstract way of defining the behaviour of this and other implementations [30, 31, 33, 36, 38]. They can also be used to define weaker isolation levels than serializability. As an illustration, we can define shard-local certification functions for snapshot isolation as follows: $f_s(T, t) = \text{COMMIT}$ iff

$$\forall x \in \text{Obj}_s. \forall v. (x, v) \in R(t) \wedge (x, \_) \in W(t) \implies (\forall t' \in T. (x, \_) \in W(t') \implies V_c(t') \leq v),$$

and $g_s(T, t) = \text{COMMIT}$ iff

$$(x, \_) \in W(t) \implies (\forall t' \in T. (x, \_) \notin W(t')).$$

The function $f_s$ restricts the global function (3) to the objects managed by the shard $s$. Since snapshot isolation allows reading stale data, the function $g_s$ only checks for write conflicts.

For shard-local certification functions to correctly approximate a given global function $f$, we require the following relationships. For a set of transactions $T \subseteq \mathcal{T}$, we write $T \mid s$ to denote the *projection* of $T$ on shard $s$, i.e., $\{t \in T \mid s \in \text{shards}(t)\}$. Then we require that

$$\forall t \in \mathcal{T}. \forall T \subseteq \mathcal{T}. f(T, t) = \text{COMMIT} \iff \forall s \in \text{shards}(t). f_s((T \mid s), t) = \text{COMMIT}. \quad (6)$$

In addition, for each shard $s$, the two functions $f_s$ and $g_s$ are required to be related to each other as follows:

$$\forall t. s \in \text{shards}(t) \implies (\forall T. g_s(T, t) = \text{COMMIT} \implies f_s(T, t) = \text{COMMIT}); \quad (7)$$

$$\forall t, t'. s \in \text{shards}(t) \cap \text{shards}(t') \implies (g_s(\{t\}, t') = \text{COMMIT} \implies f_s(\{t'\}, t) = \text{COMMIT}). \quad (8)$$

Property (7) requires the conflict check performed by $g_s$ to be no weaker than the one performed by $f_s$. Property (8) requires a form of commutativity: if $t'$ is allowed to commit after a still-pending transaction $t$, then $t$ would be allowed to commit after $t'$. The above shard-local functions for serializability and snapshot isolation satisfy (6)-(8).

**Forgetting and recalling decisions.** The protocol in Figure 1 has two additional handlers at lines 21 and 24, executed non-deterministically. As we show in §4, these are required for the abstract protocol to capture the behaviour of optimised fault-tolerant TCS implementations. Because of process crashes, such implementations may temporarily lose the information about some final decisions, and later reconstruct it from the votes at the relevant shards. In the meantime, the absence of the decisions may affect some vote computations as we explained above. The handler at line 21 forgets the decision on a transaction (but not its vote). The handler at line 24 allows processes to resend the votes they know to the coordinator, which will then resend the final decisions (line 14). This allows a process that forgot a decision to reconstruct it from the votes stored at the relevant shards.

**Correctness.** The following theorem shows the correctness of multi-shot 2PC. In particular, it shows that the shard-local concurrency control given by $f_s$ and $g_s$ correctly implements the shard-agnostic concurrency control given by a global certification function $f$.

▶ **Theorem 1.** *A transaction certification service implemented using the multi-shot 2PC protocol in Figure 1 is correct with respect to a certification function $f$, provided shard-local certification functions $f_s$ and $g_s$ satisfy (6)-(8).*

We give the proof in §A. Its main challenge is that, in multi-shot 2PC, certification orders at different shards may disagree on the order of concurrently certified transactions; however, a correct TCS has to certify transactions according to a single total order. We use the commutativity property (8) to show that per-shard certification orders arising in the protocol can be merged into the desired single total order.

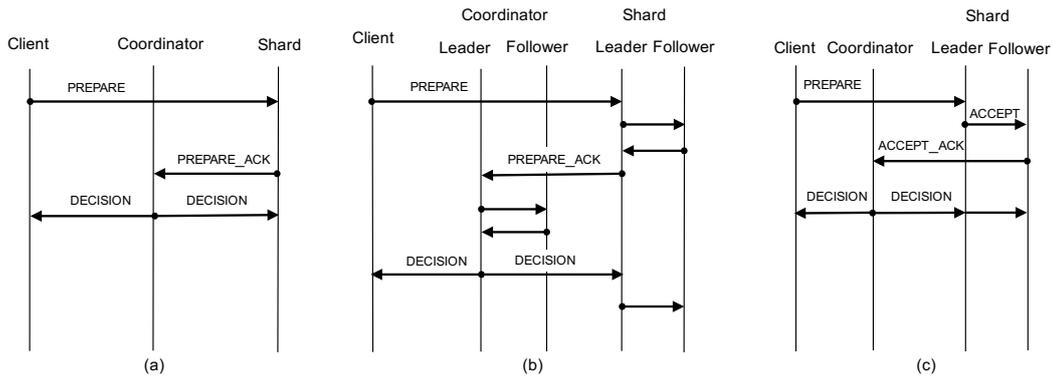

**Figure 2** Message flow diagrams illustrating the behaviour of (a) multi-shot 2PC; (b) multi-shot 2PC with shards replicated using Paxos; (c) optimised protocol weaving together multi-shot 2PC and Paxos.

## 4 Fault-Tolerant Commit Protocol

**System model.** We now weaken the assumptions of the previous section by allowing processes to fail by crashing, i.e., permanently stopping execution. We still assume that processes are connected by reliable FIFO channels in the following sense: messages are delivered in the FIFO order, and messages between non-faulty processes are guaranteed to be eventually delivered. Each shard $s$ is now managed by a group of $2f + 1$ processes, out of which at most $f$ can fail. We call a set of $f + 1$ processes in this group a *quorum* for $s$. For a shard $s$ we redefine $\mathsf{proc}(s)$ to be the set of processes managing this shard. For simplicity, we assume that the groups of processes managing different shards are disjoint.

**Vanilla protocol.** A straightforward way to implement a TCS in the above model is to use state-machine replication [35] to make a shard simulate a reliable process in multi-shot 2PC; this is usually based on a consensus protocol such as Paxos [26]. In this case, final decisions on transactions are never forgotten, and hence, the handlers at lines 21 and 24 are not simulated. Even though this approach is used by several systems [8, 13, 41], multiple researchers have observed that the resulting protocol requires an unnecessarily high number of message delays [25, 27, 40]. Namely, every action of multi-shot 2PC in Figure 2a requires an additional round trip to a quorum of processes in the same shard to persist its effect, resulting in the message-flow diagram in Figure 2b. Note that the coordinator actions have to be replicated as well, since multi-shot 2PC will block if the coordinator fails. The resulting protocol requires 7 message delays for a client to learn a decision on a transaction.

**Optimised protocol overview.** In Figures 3 and 4 we give a commit protocol that reduces the number of message delays by weaving together multi-shot 2PC across shards and a Paxos-like protocol within each shard. We omit details related to message retransmissions from the code. We illustrate the message flow of the protocol in Figure 2c and summarise the key invariants used in its proof of correctness in Figure 5.

A process maintains the same variables as in the multi-shot 2PC protocol (Figure 1) and a few additional ones. Every process in a shard is either the *leader* of the shard or a *follower*. If the leader fails, one of the followers takes over. A status variable records whether the process is a LEADER, a FOLLOWER or is in a special RECOVERING state used during leader changes. A period of time when a particular process acts as a leader is denoted using integer *ballots*. For a ballot $b \geq 1$, the process $\mathsf{leader}(b) = ((b - 1) \mod (2f + 1))$ is the leader of the ballot. At any given time, a process participates in a single ballot, stored in a variable ballot. During leader changes we also use an additional ballot variable cballot.

Unlike the vanilla protocol illustrated in Figure 2b, our protocol does not perform consensus to persist the contents of a `DECISION` message in a shard. Instead, the final decision on a transaction is sent to the members of each relevant shard asynchronously. This means that different shard members may receive the decision on a transaction at different times. Since the final decision on a transaction affects vote computations on transactions following it in the certification order (§3), computing the vote on a later transaction at different shard members may yield different outcomes. To deal with this, in our protocol only the leader constructs the certification order and computes votes. Followers are passive: they merely copy the leader's decisions. A final decision is taken into account in vote computations at a shard once it is received by the shard's leader.

**Failure-free case.** To certify a transaction $t$, a client sends it in a `PREPARE` message to the relevant shards (line 10). A process $p_i$ handles the message only when it is the leader of its shard $s_0$ (line 12). We defer the description of the cases when another process $p_j$ is resending the `PREPARE` message to $p_i$ (line 13), and when $p_i$ has already received $t$ in the past (line 14).

Upon receiving `PREPARE`($t$), the leader $p_i$ first determines a process $p$ that will serve as the coordinator of $t$. If the leader receives $t$ for the first time (line 16), then, similarly to multi-shot 2PC, it appends $t$ to the certification order and computes the vote based on the locally available information. The leader next performs an analogue of "phase 2" of Paxos, trying to convince its shard $s_0$ to accept its proposal. To this end, it sends an `ACCEPT` message to $s_0$ (including itself, for uniformity), which is analogous to the "2a" message of Paxos (line 21). The message carries the leader's ballot, the transaction $t$, its position in the certification order, the vote and the identity of $t$'s coordinator. The leader code ensures Invariant 1 in Figure 5: in a given ballot $b$, a unique transaction-vote pair can be assigned to a slot $k$ in the certification order.

A process handles an `ACCEPT` message only if it participates in the corresponding ballot (line 23). If the process has not heard about $t$ before, it stores the transaction and the vote and advances the transaction's phase to PREPARED. It then sends an `ACCEPT_ACK` message to the coordinator of $t$, analogous to the "2b" message of Paxos. This confirms that the process has accepted the transaction and the vote. The certification order at a follower is always a prefix of the certification order at the leader of the ballot the follower is in, as formalised by Invariant 2. This invariant is preserved when the follower receives `ACCEPT` messages due to the FIFO ordering of channels.

The coordinator of a transaction $t$ acts once it receives a quorum of `ACCEPT_ACK` messages for $t$ from each of its shards $s \in \mathsf{shards}(t)$, which carry the vote $d_s$ by $s$ (line 29). The coordinator computes the final decision on $t$ and sends it in `DECISION` messages to the client and to each of the relevant shards. When a process receives a decision for a transaction (line 33), the process stores it and advances the transaction's phase to DECIDED.

Once the final decision on a transaction is delivered to the leader of a shard, it is taken into account in future vote computations at this shard. Taking as an example the shard-local functions for serializability (4) and (5), if a transaction that wrote to an object $x$ is finally decided to abort, then delivering this decision to the leader may allow another transaction writing to $x$ to commit.

```
 1  next ← −1 ∈ ℤ;
 2  txn[ ] ∈ ℕ → 𝒯;
 3  vote[ ] ∈ ℕ → {COMMIT, ABORT};
 4  dec[ ] ∈ ℕ → {COMMIT, ABORT};
 5  phase[ ] ← (λk. START) ∈ ℕ → {START, PREPARED, DECIDED};
 6  status ∈ {LEADER, FOLLOWER, RECOVERING};
 7  ballot ← 0 ∈ ℕ;
 8  cballot ← 0 ∈ ℕ;

 9  function certify(t)
10  │   send PREPARE(t) to proc(shards(t));

11  when received PREPARE(t) from $p_j$ or a client
12  │   pre: status = LEADER;
13  │   if received from a process $p_j$ then $p ← p_j$ else $p ←$ coord(t);
14  │   if $\exists k. t =$ txn[k] then
15  │   │   send ACCEPT(ballot, k, t, vote[k], p) to proc($s_0$)
16  │   else
17  │   │   next ← next + 1;
18  │   │   txn[next] ← t;
19  │   │   vote[next] ← $f_{s_0}$({txn[k] | k < next ∧ phase[k] = DECIDED ∧ dec[k] = COMMIT}, t) ⊓
                        $g_{s_0}$({txn[k] | k < next ∧ phase[k] = PREPARED ∧ vote[k] = COMMIT}, t);
20  │   │   phase[next] ← PREPARED;
21  │   │   send ACCEPT(ballot, next, t, vote[next], p) to $s_0$;

22  when received ACCEPT(b, k, t, d, p)
23  │   pre: status ∈ {LEADER, FOLLOWER} ∧ ballot = b;
24  │   if phase[k] = START then
25  │   │   txn[k] ← t;
26  │   │   vote[k] ← d;
27  │   │   phase[k] ← PREPARED;
28  │   send ACCEPT_ACK($s_0$, b, k, t, d) to p;

29  when for every $s ∈$ shards(t) received a quorum of
    ACCEPT_ACK(s, $b_g$, $pos_g$, t, $d_g$)
30  │   send DECISION(t, $\bigsqcap_{s \in \text{shards}(t)} d_s$) to client(t);
31  │   forall $s ∈$ shards(t) do
32  │   │   send DECISION($b_s$, $pos_s$, $\bigsqcap_{s \in \text{shards}(t)} d_s$) to proc(s)

33  when received DECISION(b, k, d)
34  │   pre: status ∈ {LEADER, FOLLOWER} ∧ ballot ≥ b ∧ phase[k] = PREPARED;
35  │   dec[k] ← d;
36  │   phase[k] ← DECIDED;
```

**Figure 3** Fault-tolerant commit protocol at a process $p_i$ in a shard $s_0$: failure-free case.

```
37  function recover()
38  |   send NEW_LEADER(any ballot b such that b > ballot ∧ leader(ballot) = p_i) to s_0;

39  when received NEW_LEADER(b) from p_j
40  |   pre: b > ballot;
41  |   status ← RECOVERING;
42  |   ballot ← b;
43  |   send NEW_LEADER_ACK(ballot, cballot, txn, vote, dec, phase) to p_j;

44  when received {NEW_LEADER_ACK(b, cballot_j, txn_j, vote_j, dec_j, phase_j) | p_j ∈ Q}
         from a quorum Q in s_0
45  |   pre: status = RECOVERING ∧ ballot = b;
46  |   var J ← the set of j with the maximal cballot_j;
47  |   forall k do
48  |       if ∃j ∈ J. phase_j[k] ≥ PREPARED then
49  |           txn[k] ← txn_j[k];
50  |           vote[k] ← vote_j[k];
51  |           phase[k] ← PREPARED;
52  |       if ∃j. phase_j[k] = DECIDED then
53  |           dec ← dec_j[k];
54  |           phase[k] ← DECIDED;
55  |   next ← min{k | phase[k] ≠ START};
56  |   cballot ← b;
57  |   status ← LEADER;
58  |   send NEW_STATE(b, txn, vote, dec, phase) to proc(s_0) \ {p_i};

59  when received NEW_STATE(b, txn, vote, dec, phase) from p_j
60  |   pre: b ≥ ballot;
61  |   status ← FOLLOWER;
62  |   cballot ← b;
63  |   txn ← txn;
64  |   vote ← vote;
65  |   dec ← dec;
66  |   phase ← phase;

67  function retry(k)
68  |   pre: phase[k] = PREPARED;
69  |   send PREPARE(txn[k]) to proc(shards(txn[k]));
```

**Figure 4** Fault-tolerant commit protocol at a process $p_i$ in a shard $s_0$: recovery.

**Leader recovery.** We next explain how the protocol deals with failures, starting from a leader failure. The goal of the leader recovery procedure is to preserve Invariant 3: if in a ballot $b$ a shard $s$ accepted a vote $d$ on a transaction $t$ at the position $k$ in the certification order, then this vote will persist in all future ballots; this is furthermore true for all votes the leader of ballot $b$ took into account when computing $d$. The latter property is necessary for the shard to simulate the behaviour of a reliable process in multi-shot 2PC that maintains a unique certification order. To ensure this property, our recovery procedure includes an additional message from the new leader to the followers ensuring that, before a follower starts accepting proposals from the new leader, it has brought its state in sync with that of the leader (this is similar to [21, 28]). The ballot of the last leader a follower synchronised with in this way is recorded in cballot.

We now describe the recovery procedure in detail. When a process $p_i$ suspects the leader of its shard of failure, it may try to become a new leader by executing the `recover` function (line 37). The process picks a ballot that it leads higher than ballot and sends it in a `NEW_LEADER` message to the shard members (including itself); this message is analogous to the "1a" message in Paxos. When a process receives a `NEW_LEADER`($b$) message (line 39), it first checks that the proposed ballot $b$ is higher than his. In this case, it sets its ballot to $b$ and changes its status to RECOVERING, which causes it to stop processing `PREPARE`, `ACCEPT` and `DECISION` messages. It then replies to the new leader with a `NEW_LEADER_ACK` message containing all components of its state, analogous to the "1b" message of Paxos.

The new leader waits until it receives `NEW_LEADER_ACK` messages from a quorum of shard members (line 44). Based on the states reported by the processes, it computes a new state from which to start certifying transactions. Like in Paxos, the leader focusses on the states of processes that reported the maximal cballot (line 46): if the $k$-th transaction is PREPARED at such a process, then the leader marks it as accepted and copies the vote; furthermore, if the transaction is DECIDED at some process (with any ballot number), then the leader marks it as decided and copies the final decision. Given Invariant 2, we can show that the resulting certification order does not have holes: if a transaction is PREPARED or DECIDED, then so are the previous transactions in the certification order.

The leader sets next to the length of the merged sequence of transactions, cballot to the new ballot and status to LEADER, which allows it to start processing new transactions (lines 55-57). It then sends a `NEW_STATE` message to other shard members, containing the new state (line 58). Upon receiving this message (line 59), a process overwrites its state with the one provided, changes its status to FOLLOWER, and sets cballot to $b$, thereby recording the fact that it has synchronised with the leader of $b$. Note that the process will not accept transactions from the new leader until it receives the `NEW_STATE` message. This ensures that Invariant 2 is preserved when the process receives the first `ACCEPT` message in the new ballot.

**Coordinator recovery.** If a process that accepted a transaction $t$ does not receive the final decision on it, this may be because the coordinator of $t$ has failed. In this case the process may decide to become a new coordinator by executing the `retry` function (line 67). For this the process just re-sends the `PREPARE`($t$) message to the shards of $t$. A leader handles the `PREPARE`($t$) message received from another process $p_j$ similarly to one received from a client. If it has already certified the transaction $t$, it re-sends the corresponding `ACCEPT` message to the shard members, asking them to reply to $p_j$ (line 14). Otherwise, it handles $t$ as before. In the end, a quorum of processes in each shard will reply to the new coordinator (line 28), which will then broadcast the final decision (lines 30-31). Note that the check at line 14 ensures Invariants 4 and 5: in a given ballot $b$, a transaction $t$ can only be assigned to a

1. If $\texttt{ACCEPT}(b, k, t_1, d_1, \_)$ and $\texttt{ACCEPT}(b, k, t_2, d_2, \_)$ messages are sent to the same shard, then $t_1 = t_2$ and $d_1 = d_2$.
2. After a process receives and acknowledges $\texttt{ACCEPT}(b, k, t, d, \_)$, we have $\textsf{txn} = txn\!\downarrow_k$ and $\textsf{vote} = vote\!\downarrow_k$, where $txn$ and $vote$ are the values of the arrays $\textsf{txn}$ and $\textsf{vote}$ at $\textsf{leader}(b)$ when it sent the $\texttt{ACCEPT}$ message.
3. Assume that a quorum of processes in $s$ received $\texttt{ACCEPT}(b, k, t, d, \_)$ and responded to it with $\texttt{ACCEPT\_ACK}(s, b, k, t, d)$, and at the time $\textsf{leader}(b)$ sent $\texttt{ACCEPT}(b, k, t, d, \_)$ it had $\textsf{txn}\!\downarrow_k = txn$ and $\textsf{vote}\!\downarrow_k = vote$. Whenever at a process in $s$ we have $\textsf{status} \in \{\textsc{leader}, \textsc{follower}\}$ and $\textsf{ballot} = b' > b$, we also have $\textsf{txn}\!\downarrow_k = txn$ and $\textsf{vote}\!\downarrow_k = vote$.
4. If $\texttt{ACCEPT}(b, k_1, t, \_, \_)$ and $\texttt{ACCEPT}(b, k_2, t, \_, \_)$ messages are sent to the same shard, then $k_1 = k_2$.
5. At any process, all transactions in the $\textsf{txn}$ array are distinct.
6. a. For any messages $\texttt{DECISION}(\_, k, d_1)$ and $\texttt{DECISION}(\_, k, d_2)$ sent to processes in the same shard, we have $d_1 = d_2$.
    b. For any messages $\texttt{DECISION}(t, d_1)$ and $\texttt{DECISION}(t, d_2)$ sent, we have $d_1 = d_2$.
7. a. Assume that a quorum of processes in $s$ have sent $\texttt{ACCEPT\_ACK}(s, b_1, k, t_1, d_1)$ and a quorum of processes in $s$ have sent $\texttt{ACCEPT\_ACK}(s, b_2, k, t_2, d_2)$. Then $t_1 = t_2$ and $d_1 = d_2$.
    b. Assume that a quorum of processes in $s$ have sent $\texttt{ACCEPT\_ACK}(s, b_1, k_1, t, d_1)$ and a quorum of processes in $s$ have sent $\texttt{ACCEPT\_ACK}(s, b_2, k_2, t, d_2)$. Then $k_1 = k_2$ and $d_1 = d_2$.

**Figure 5** Key invariants of the fault-tolerant protocol. We let $\alpha\!\downarrow_k$ be the prefix of the sequence $\alpha$ of length $k$.

single slot in the certification order, and all transactions in the $\textsf{txn}$ array are distinct.

Our protocol allows any number of processes to become coordinators of a transaction at the same time: unlike in the vanilla protocol of Figure 2b, coordinators are not consistently replicated. Nevertheless, the protocol ensures that they will all reach the same decision, even in case of leader changes. We formalise this in Invariant 6: part (a) ensures an agreement on the decision on the $k$-th transaction in the certification order at a given shard; part (b) ensures a system-wide agreement on the decision on a given transaction $t$. The latter part establishes that the fault-tolerant protocol computes a unique decision on each transaction.

By the structure of the hander at line 29, Invariant 6 follows from Invariant 7, since, if a coordinator has computed the final decision on a transaction, then a quorum of processes in each relevant shard has accepted a corresponding vote. We next prove Invariant 7.

**Proof of Invariant 7.** *(a)* Assume that quorums of processes in $s$ have sent $\texttt{ACCEPT\_ACK}(s, b_1, k, t_1, d_1)$ and $\texttt{ACCEPT\_ACK}(s, b_2, k, t_2, d_2)$. Then $\texttt{ACCEPT}(b_1, k, t_1, d_1, \_)$ and $\texttt{ACCEPT}(b_2, k, t_2, d_2, \_)$ have been sent to $s$. Assume without loss of generality that $b_1 \leq b_2$. If $b_1 = b_2$, then by Invariant 1 we must have $t_1 = t_2$ and $d_1 = d_2$. Assume now that $b_1 < b_2$. By Invariant 3, when $\textsf{leader}(b_2)$ sends the $\texttt{ACCEPT}$ message, it has $\textsf{txn}[k] = t_1$. But then due to the check at line 14, we again must have $t_1 = t_2$ and $d_1 = d_2$.

*(b)* Assume that quorums of processes in $s$ have sent $\texttt{ACCEPT\_ACK}(s, b_1, k_1, t, d_1)$ and $\texttt{ACCEPT\_ACK}(s, b_2, k_2, t, d_2)$. Then $\texttt{ACCEPT}(b_1, k_1, t, d_1, \_)$ and $\texttt{ACCEPT}(b_2, k_2, t, d_2, \_)$ have been sent to $s$. Without loss of generality, we can assume $b_1 \leq b_2$. We first show that $k_1 = k_2$. If $b_1 = b_2$, then we must have $k_1 = k_2$ by Invariant 4. Assume now that $b_1 < b_2$. By Invariant 3, when $\textsf{leader}(b_2)$ sends the $\texttt{ACCEPT}$ message, it has $\textsf{txn}[k_1] = t$. But then due to the check at line 14 and Invariant 5, we again must have $k_1 = k_2$. Hence, $k_1 = k_2$. But then by Invariant 7a we must also have $d_1 = d_2$. □

**Protocol correctness.** We only establish the safety of the protocol (in the sense of the correctness condition in §2) and leave guaranteeing liveness to standard means, such as assuming either an oracle that is eventually able to elect a consistent leader in every shard [5], or that the system eventually behaves synchronously for sufficiently long [11].

▶ **Theorem 2.** *The fault-tolerant commit protocol in Figures 3-4 simulates the multi-shot 2PC protocol in Figure 1.*

We give the proof in §B. Its main idea is to show that, in an execution of the fault-tolerant protocol, each shard produces a single certification order on transactions from which votes and final decisions are computed. These certification orders determine the desired execution of the multi-shot 2PC protocol. We prove the existence of a single per-shard certification order using Invariant 3, showing that certification orders and votes used to compute decisions persist across leader changes. However, this property does not hold of final decisions, and it is this feature that necessitates adding transitions for forgetting and recalling final decisions to the protocol in Figure 1 (lines 21 and 24).

For example, assume that the leader of a ballot $b$ at a shard $s$ receives the decision ABORT on a transaction $t$. The leader will then take this decision into account in its vote computations, e.g., allowing transactions conflicting with $t$ to commit. However, if the leader fails, a new leader may not find out about the final decision on $t$ if this decision has not yet reached other shard members. This leader will not be able to take the decision into account in its vote computations until it reconstructs the decision from the votes at the relevant shards (line 67). Forgetting and recalling the final decisions in the multi-shot 2PC protocol captures how such scenarios affect vote computations.

**Optimisations.** Our protocol allows the client and the relevant servers to learn the decision on a transaction in four message delays, including communication with the client (Figure 2c). As in standard Paxos, this can be further reduced to three message delays at the expense of increasing the number of messages sent by eliminating the coordinator: processes can send their `ACCEPT_ACK` messages for a transaction directly to all processes in the relevant shards and to the client. Each process can then compute the final decision independently. The resulting time complexity matches the lower bounds for consensus [6, 23] and non-blocking atomic commit [12].

In practice, the computation of a shard-local function for $s$ depends only on the objects managed by $s$: e.g., $\mathsf{Obj}_s$ for (4) and (5). Hence, once a process at a shard $s$ receives the final decision on a transaction $t$, it may discard the data of $t$ irrelevant to $s$. Note that the same cannot be done when $t$ is only prepared, since the complete information about it may be needed to recover from coordinator failure (line 67).

## 5   Related Work

The existing work on the Atomic Commit Problem (ACP) treats it as a one-shot problem with the votes being provided as the problem inputs. The classic ACP solution is the Two-Phase Commit (2PC) protocol [14], which blocks in the event of the coordinator failure. The *non-blocking* variant of ACP known as Non-Blocking Atomic Commit (NBAC) [37] has been extensively studied in both the distributed computing and database communities [12, 15, 16, 17, 19, 22, 32, 37]. The Three-Phase Commit (3PC) family of protocols [2, 3, 12, 22, 37] solve

NBAC by augmenting 2PC with an extra message exchange round in the failure-free case. Paxos Commit [15] and Guerraoui et al. [17] avoid extra message delays by instead replicating the 2PC participants through consensus instances. While our fault-tolerant protocol builds upon similar ideas to optimise the number of failure-free message delays, it nonetheless solves a more general problem (TCS) by requiring the output decisions to be compatible with the given isolation level.

Recently, Guerraoui and Wang [18] have systematically studied the failure-free complexity of NBAC (in terms of both message delays and number of messages) for various combinations of the correctness properties and failure models. The complexity of certifying a transaction in the failure-free runs of our crash fault-tolerant TCS implementation (provided the coordinator is replaced with all-to-all communication) matches the tight bounds for the most robust version of NBAC considered in [18], which suggests it is optimal. A comprehensive study of the TCS complexity in the absence of failures is the subject of future work.

Our multi-shot 2PC protocol is inspired by how 2PC is used in a number of systems [8, 13, 30, 31, 33, 36, 38]. Unlike prior works, we formalise how 2PC interacts with concurrency control in such systems in a way that is parametric in the isolation level provided and give conditions for its correctness, i.e., (6)-(8). A number of systems based on deferred update replication [29] used non-fault-tolerant 2PC for transaction commit [30, 31, 33, 38]. Our formalisation of the TCS problem should allow making them fault-tolerant using protocols of the kind we presented in §4.

Multiple researchers have observed that implementing transaction commit by layering 2PC on top of Paxos is suboptimal and proposed possible solutions [10, 25, 27, 36, 40]. In comparison to our work, they did not formulate a stand-alone certification problem, but integrated certification with the overall transaction processing protocol for a particular isolation level and corresponding optimisations.

In more detail, Kraska et al. [25] and Zhang et al. [40] presented sharded transaction processing systems, respectively called MDCC and TAPIR, that aim to minimise the latency of transaction commit in a geo-distributed setting. The protocols used are leaderless: to compute the vote, the coordinator of a transaction contacts processes in each relevant shard directly; if there is a disagreement between the votes computed by different processes, additional message exchanges are needed to resolve it. This makes the worse-case failure-free time complexity of the protocols higher than that of our fault-tolerant protocol. The protocols were formulated for particular isolation levels (a variant of Read Committed in MDCC and serializability in TAPIR). Both MDCC and TAPIR are significantly more complex than our fault-tolerant commit protocol and lack rigorous proofs of correctness.

Sciascia et al. proposed Scalable Deferred Update Replication [36] for implementing serializable transactions in sharded systems. Like the vanilla protocol in §4, their protocol keeps shards consistent using black-box consensus. It avoids executing consensus to persist a final decision by just not taking final decisions into account in vote computations. This solution, specific to their conflict check for serializability, is suboptimal: if a prepared transaction $t$ aborts, it will still cause conflicting transactions to abort until their read timestamp goes above the write timestamp of $t$.

Dragojević et al. presented a FaRM transactional processing system based on RDMA [10]. Like in our fault-tolerant protocol, in the FaRM atomic commit protocol only shard leaders compute certification votes. However, recovery in FaRM is simplified by the use of leases and an external reconfiguration engine.

Mahmoud et al. proposed Replicated Commit [27], which reduces the latency of transac-tion commit by layering Paxos on top of 2PC, instead of the other way round. This approach relies on 2PC deciding ABORT only in case of failures, but not because of concurrency control. This requires integrating the transaction commit protocol with two-phase locking and does not allow using it with optimistic concurrency control.

Schiper et al. proposed an alternative approach to implementing deferred update replication in sharded systems [34]. This distributes transactions to shards for certification using genuine atomic multicast [9], which avoids the need for a separate fault-tolerant commit protocol. However, atomic multicast is more expensive than consensus: the best known implementation requires 4 message delays to deliver a message, in addition to a varying convoy effect among different transactions [7]. The resulting overall latency of certification is 5 message delays plus the convoy effect.

Our fault-tolerant protocol follows the primary/backup state machine replication approach in imposing the leader order on transactions certified within each shard. This is inspired by the design of some total order broadcast protocols, such as Zab [21] and Viewstamped Replication [28]. Kokocinski et al. [24] have previously explored the idea of delegating the certification decision to a single leader in the context of deferred update replication. However, they only considered a non-sharded setting, and did not provide full implementation details and a correctness proof. In particular, it is unclear how correctness is maintained under leader changes in their protocol.

## 6  Conclusion

In this paper we have made the first step towards building a theory of distributed transaction commit in modern transaction processing systems, which captures interactions between atomic commit and concurrency control. We proposed a new problem of transaction certification service and an abstract protocol solving it among reliable processes. From this, we have systematically derived a provably correct optimised fault-tolerant protocol.

For conciseness, in this paper we focussed on transaction processing systems using optimistic concurrency control. We hope that, in the future, our framework can be generalised to systems that employ pessimistic concurrency control or a mixture of the two. The simple and leader-driven nature of our optimised protocol should also allow porting it to the Byzantine fault-tolerant setting by integrating ideas from consensus protocols such as PBFT [4].

# APPENDIX

## A  Proof of Theorem 1

Fix an execution of the protocol producing a history $h$. We start by establishing some constraints that the votes and decisions computed in this execution satisfy (Figure 6). Let $T$ be the set of transactions $t$ such that $\texttt{certify}(t)$ occurs in $h$. For some of transactions $t \in T$ and shards $s \in \textsf{shards}(t)$, we define the certification order position $pos_s[t]$ and a vote $d_s[t]$ computed by the protocol. These are defined as follows:

> Consider $t \in T$ and $s \in \textsf{shards}(t)$. Assume the process $\textsf{proc}(s)$ sent $\texttt{PREPARE\_ACK}(s, k, t, d)$ and right after this it had $\textsf{txn} = txn$ and $\textsf{vote} = vote$. Then for every $j \leq k$ we let $pos_s[txn[j]] = j$ and $d_s[txn[j]] = vote[j]$.

It is easy to see that this defines $pos_s[t]$ and $d_s[t]$ uniquely. Furthermore, by the structure of the hander at line 14 in Figure 1, for each $t$ such that $\texttt{decide}(t, d[t])$ occurs in $h$, $d_s[t]$ is defined for all $s \in \textsf{shards}(t)$ and (11) holds. It is also easy to check that the order determined by $pos_s$ is downclosed (12), transactions are assigned to distinct positions by $pos_s$ (13), and these positions are consistent with the real-time ordering of operations (14). We can furthermore establish that vote computations in the optimised commit protocol satisfy the constraints (15)-(17) for some sets of transactions $T_s[t]$ and $P_s[t]$.

For each action $\texttt{decide}(\_, \textsc{commit}) \in \textsf{act}(h)$ in $h$, we assign it a linearization point at the time when the corresponding $\texttt{DECISION}$ message is sent to the client for the first time. To prove the theorem, it is enough to show that the induced linearization of $h \mid \textsf{committed}(h)$ is legal with respect to $f$. We do this by induction on its length $k \geq 0$. This trivially holds when $k = 0$. Assume now that the linearization $\ell_k$ of length $k$ is legal and consider a linearization $\ell_{k+1}$ of length $k + 1$.

For all $i = 1..(k+1)$, let $\texttt{decide}(t_i, \textsc{commit})$ be the $i$-th $\texttt{decide}$ action in $\ell_k$. Let $T_0 = \emptyset$, and for all $j = 1..k$ let $T_j = T_{j-1} \cup \{t_j\}$. By the induction hypothesis, $f(T_{i-1}, t_i) = \textsc{commit}$ for all $i = 1..k$. We show that $f(T_k, t_{k+1}) = \textsc{commit}$, which implies that $\ell_{k+1}$ is legal with respect to $f$.

Let
$$T = \{t_i \mid \neg(\texttt{decide}(t_i, \_) \prec_h \texttt{certify}(t_{k+1})) \land 1 \leq i \leq k\}, \quad T' = T_k \setminus T.$$

Fix a shard $s \in \textsf{shards}(t_{k+1})$. Let
$$U_0 = \{t' \mid pos_s[t'] < pos_s[t_{k+1}]\} \cap (T_k \mid s).$$

Then
$$U_0 \subseteq \{t' \mid pos_s[t'] < pos_s[t_{k+1}] \land d[t'] = \textsc{commit}\}. \tag{9}$$

We have $d[t_{k+1}] = \textsc{commit}$, which by (11) implies $d_s[t_{k+1}] = \textsc{commit}$. Then by (15) we have
$$f_s(T_s[t_{k+1}], t_{k+1}) \sqcap g_s(P_s[t_{k+1}], t_{k+1}) = d_s[t_{k+1}] = \textsc{commit},$$

From (9), (16) and (17), we get $U_0 \subseteq T_s[t_{k+1}] \cup P_s[t_{k+1}]$. Then by (7) and (1), we have
$$f_s(U_0, t_{k+1}) = \textsc{commit}. \tag{10}$$

By (14), $(T' \mid s) \subseteq U_0$. We now augment $U_0$ with the transactions in $(T \mid s) \setminus U_0$ as follows. Let $t^1, \ldots, t^m$ where $m = |(T \mid s) \setminus U_0|$ be the set of transactions in $(T \mid s) \setminus U_0$ sorted according to $pos_s$. Consider a sequence of sets $U_0, \ldots, U_m$ such that $U_i = U_{i-1} \cup \{t^i\}$ for $i = 1..m$. We now prove that $f_s(U_i, t_{k+1}) = \textsc{commit}$ for $i = 0..m$ by induction on $i$.

$$d[t] = \bigsqcap \{d_s[t] \mid s \in \mathsf{shards}(t)\} \tag{11}$$

$$\forall t. \, (pos_s[t] \text{ is defined}) \implies \forall j < pos_s[t]. \, \exists t'. \, pos_s[t'] = j \tag{12}$$

$$\forall t_1, t_2. \, t_1 \neq t_2 \implies pos_s[t_1] \neq pos_s[t_2] \tag{13}$$

$$\forall t, t', s. \, \mathtt{decide}(t, \_) \prec_h \mathtt{certify}(t') \wedge s \in \mathsf{shards}(t) \cap \mathsf{shards}(t') \implies pos_s[t] < pos_s[t'] \tag{14}$$

$$d_s[t] = f_s(T_s[t], t) \sqcap g_s(P_s[t], t) \tag{15}$$

$$T_s[t] = \{t' \mid pos_s[t'] < pos_s[t] \wedge d[t'] = \textsc{commit}\} \setminus P_s[t] \tag{16}$$

$$P_s[t] \subseteq \{t' \mid pos_s[t'] < pos_s[t] \wedge d_s[t'] = \textsc{commit}\} \tag{17}$$

**Figure 6** Constraints on the votes computed by the multi-shot 2PC protocol.

First, note that (10) implies that the claim is true for $i = 0$. Next, assume that $f_s(U_j, t_{k+1}) = \textsc{commit}$ for all $j = 0..(i-1)$, and consider $f_s(U_i, t_{k+1})$ where $U_i = U_{i-1} \cup \{t^i\}$. Since $t^i$ comes before $t_{k+1}$ in the linearization, no message $\mathtt{DECISION}(t_{k+1}, \_)$ has been sent when shard $s$ votes on $t^i$. Hence, we must have $t_{k+1} \in P_s[t^i]$ and, by (1), $g_s(\{t_{k+1}\}, t^i) = \textsc{commit}$. By (8), this implies $f_s(\{t^i\}, t_{k+1}) = \textsc{commit}$. From this and the induction hypothesis, by (1) we get

$$f_s(U_i, t_{k+1}) = f_s(U_{i-1} \cup \{t^i\}, t_{k+1}) = \textsc{commit},$$

which concludes the proof of the induction step.

We have established $f_s(U_m, t_{k+1}) = \textsc{commit}$. Since $(T_k \mid s) = ((T \cup T') \mid s) = U_m$, this implies $f_s((T_k \mid s), t_{k+1}) = \textsc{commit}$. Thus, for all $s \in \mathsf{shards}(t_{k+1})$ we have $f_s((T_k \mid s), t_{k+1}) = \textsc{commit}$, which by (6) implies $f(T_k, t_{k+1}) = \textsc{commit}$, as required. □

## B  Proof of Theorem 2

▶ **Proposition 3.** *The fault-tolerant commit protocol satisfies the following invariants:*
1. *If* $\mathsf{txn}[k]$ *is defined, then* $\mathsf{phase}[k] \geq \textsc{prepared}$.
2. *If at a process we have* $\mathsf{status} \in \{\textsc{follower}, \textsc{leader}\}$, *then* $\mathsf{ballot} = \mathsf{cballot}$.

**Proof of Invariant 3.** We prove the invariant by induction on $b'$. Assume that the invariant holds for all $b' < b''$. We now show it for $b' = b''$. Assume that at some point a quorum $Q$ of processes in $s$ have received $\mathtt{ACCEPT}(b, k, t, d, \_)$ and responded to it with $\mathtt{ACCEPT\_ACK}(s, b, k, t, d)$, and at the time $\mathsf{leader}(b)$ sent $\mathtt{ACCEPT}(b, k, t, d, \_)$ it had $\mathsf{txn}\!\downarrow_k = txn$ and $\mathsf{vote}\!\downarrow_k = vote$.

Consider the transition by a process $p_i$ in line 44, which handles the receipt of $\mathtt{NEW\_LEADER\_ACK}$ messages and upon which $p_i$ sets $\mathsf{status} = \textsc{leader}$. Assume that after this $\mathsf{ballot} = b'' > b$ at the process. We show that we also have $\mathsf{txn}\!\downarrow_k = txn$ and $\mathsf{vote}\!\downarrow_k = vote$. Assume $p_i$ received messages

$$\mathtt{NEW\_LEADER\_ACK}(b'', cballot_j, txn_j, vote_j, dec_j, phase_j)$$

from a quorum $Q'$ of processes $p_j$. Let $b_0 = \max\{cballot_j \mid p_j \in Q'\}$ and $J = \{j \mid cballot_j = b_0\}$. Then $b_0 < b''$.

# APPENDIX

We have $Q \cap Q' \neq \emptyset$. Thus, some process process $p_{j_0}$ must have sent both `ACCEPT_ACK` and `NEW_LEADER_ACK` messages. Then $p_{j_0}$ must have sent its `ACCEPT_ACK` message before the `NEW_LEADER_ACK` message. By the precondition in line 23 and Proposition 3(2), when $p_{j_0}$ sent the `ACCEPT_ACK` message, it must have had cballot $= b$. Then when $p_{j_0}$ sent its `NEW_LEADER_ACK` message, it had cballot $\geq b$. Hence, $cballot_{j_0} \geq b$ and $b_0 \geq b$, so that $b_0 \neq 0$ and $J \neq \emptyset$.

Consider first the case when $b_0 = b$. Since $p_{j_0} \in Q'$ has received and acknowledged $\texttt{ACCEPT}(b, k, t, d, \_)$, by Invariant 2 we have $txn = txn_{j_0}\!\downarrow_k$ and $vote = vote_{j_0}\!\downarrow_k$. Furthermore, for any other process $p_j$ such that $j \in J$, if $phase_j[k] \geq \text{PREPARED}$, then $txn_{j_0}\!\downarrow_k$ is a prefix of $txn$ and $vote_{j_0}\!\downarrow_k$ is a prefix of $vote$. By Proposition 3(1) we also have $\forall l \leq k.\, phase_{j_0}[l] \geq \text{PREPARED}$. Hence, right after $p_i$ executes the transition in line 44, we have $txn = \textsf{txn}\!\downarrow_k$ and $vote = \textsf{vote}\!\downarrow_k$, as required.

Assume now that $b_0 > b$. For some process $p_j \in Q'$ we have $cballot_j = b_0$. Since $b_0 \neq 0$, either $p_j = \textsf{leader}(b_0)$ or $p_j$ must have received a `NEW_STATE` message from $\textsf{leader}(b_0)$. After the transition of $\textsf{leader}(b_0)$ in line 44 sending `NEW_STATE` messages, at $\textsf{leader}(b_0)$ we have status $= \text{LEADER}$ and ballot $= b_0 > b$. By induction hypothesis, $\textsf{leader}(b_0)$ had $txn = \textsf{txn}\!\downarrow_k$ and $vote = \textsf{vote}\!\downarrow_k$ right after executing the transition in line 44. Hence, the messages $\texttt{NEW\_STATE}(b_0, txn', vote', dec', phase')$ it sends in this transition are such that $txn = txn'\!\downarrow_k$ and $vote = vote'\!\downarrow_k$. Then all processes $p_j$ with $cballot_j = b_0$ are either such that $p_j = \textsf{leader}(b_0)$ or have received and acknowledged the `NEW_STATE` message from $\textsf{leader}(b_0)$. After each such process sends `NEW_STATE` messages or receives them, at this process we have $txn = \textsf{txn}\!\downarrow_k$ and $vote = \textsf{vote}\!\downarrow_k$. From this point until the process receives a `NEW_LEADER` message from $p_i$, $\textsf{txn}\!\downarrow_k$ and $\textsf{vote}\!\downarrow_k$ are unchanged. Hence, for each $j \in J$ we have $txn = txn_j\!\downarrow_k$ and $vote = vote_j\!\downarrow_k$. By Proposition 3(1) we also have $\forall l \leq k.\, phase_j[l] \geq \text{PREPARED}$. Then right after $p_i$ executes the transition in line 44, we have $txn = \textsf{txn}\!\downarrow_k$ and $vote = \textsf{vote}\!\downarrow_k$, as required.

Consider now the transition by a process $p_i$ in line 59, which handles the receipt of a message $\texttt{NEW\_STATE}(b, txn', vote', dec', phase')$ from a process $p_j$ and upon which $p_i$ sets status $= \text{FOLLOWER}$. Assume that after this ballot $= b'' > b$ at the process. We show that we also have $\textsf{txn}\!\downarrow_k = txn$ and $\textsf{vote}\!\downarrow_k = vote$. The process $p_j$ has status $= \text{LEADER}$ after sending the `NEW_STATE` message. Above we established that $p_j$ has $txn = \textsf{txn}\!\downarrow_k$ and $vote = \textsf{vote}\!\downarrow_k$ right after sending the `NEW_STATE` message. Hence, $txn = txn'\!\downarrow_k$ and $vote = vote'\!\downarrow_k$. Then after $p_i$ executes the transition in line 59, we have $txn = \textsf{txn}\!\downarrow_k$ and $vote = \textsf{vote}\!\downarrow_k$, as required. $\square$

▶ **Proposition 4.** **(a)** *If at a process in a shard $s$ we have* ballot $= b'$, phase$[k] = \text{DECIDED}$ *and* dec$[k] = d$, *then a* $\texttt{DECISION}(\_, k, d)$ *message has been sent to $s$, where $b \leq b'$.*
**(b)** *If at a process we have* phase$[k] = \text{DECIDED}$ *and* dec$[k] = \text{COMMIT}$, *then* vote$[k] = \text{COMMIT}$.

**Proof.** *(a)*. We prove the invariant by induction on the length of the protocol execution. The transitions that may set phase$[k] = \text{DECIDED}$ and thereby affect the validity of the invariant are those at lines 33, 44 and 59. The transition at line 33 trivially preserves the invariant. If a transition at lines 44 or 59 is executed and sets phase$[k] = \text{DECIDED}$, then the premiss of the invariant must have held earlier at some process with ballot $\leq b'$. Then the required follows from the induction hypothesis.

*(b)*. Follows from item (a) and Invariant 3. $\square$

**Proof of Theorem 2.** Fix a finite execution of the fault-tolerant commit protocol with a history $h$. We show that there exists an execution of the multi-shot 2PC protocol with the same history. Let $T$ be the set of transactions $t$ such that $\texttt{certify}(t)$ occurs in $h$. For some of transactions $t \in T$ and shards $s \in \mathsf{shards}(t)$, we define the certification order position $pos_s[t]$ and a vote $d_s[t]$ computed by the protocol as follows:

Consider $t \in T$ and $s \in \mathsf{shards}(t)$. Assume that a quorum of processes in $s$ received $\texttt{ACCEPT}(b, k, t, d, \_)$ and responded to it with $\texttt{ACCEPT\_ACK}(s, b, k, t, d)$, and at the time $\mathsf{leader}(b)$ sent $\texttt{ACCEPT}(b, k, t, d, \_)$ it had $\mathsf{txn}\!\downarrow_k = txn$ and $\mathsf{vote}\!\downarrow_k = vote$. Then for every $j \leq k$ we let $pos_s[\mathsf{txn}[j]] = j$ and $d_s[\mathsf{txn}[j]] = vote[j]$.

According to Invariants 3 and 5, this defines $pos_s[t]$ and $d_s[t]$ uniquely and (13) in Figure 6 holds. Furthermore, by the structure of the hander at line 29, for each $t$ such that $\texttt{decide}(t, d[t])$ occurs in $h$, $d_s[t]$ is defined for all $s \in \mathsf{shards}(t)$ and (11) holds. By construction, (12) holds.

We now prove (14). Consider $t, t', s$ such that

$$\texttt{decide}(t, \_) \prec_h \texttt{certify}(t') \wedge s \in \mathsf{shards}(t) \cap \mathsf{shards}(t').$$

Let $\texttt{DECISION}(b, pos_s[t], \_)$ be the message send to the shard $s$ when the $\texttt{decide}(t, \_)$ action was generated. Let $b'$ be some ballot at which $pos_s[t']$ is defined according to the above definition. Assume first that $b' < b$. Then by Invariant 3 when the leader of $b$ starts operating, it has $\mathsf{txn}[pos_s[t']] = t'$. But then $\texttt{certify}(t')$ must have occurred before $\texttt{decide}(t, \_)$. Hence, $b \leq b'$. By Invariant 3 when the leader of $b'$ receives $\texttt{PREPARE}(t')$, it has $\mathsf{txn}[pos_s[t]] = t$. But then $pos_s[t] < pos_s[t']$, which proves (14).

We prove (15)-(17) using the following proposition.

▶ **Proposition 5.** *The following always holds at any process in a shard $s$:*

$\forall k. (\mathsf{vote}[k] \text{ is defined}) \implies \exists T, P. \mathsf{vote}[k] = f_s(T, \mathsf{txn}[k]) \sqcap g_s(P, \mathsf{txn}[k]) \wedge$
$T = \{\mathsf{txn}[k'] \mid k' < k \wedge \mathsf{vote}[k'] = \textsc{commit} \wedge d[\mathsf{txn}[k']] = \textsc{commit}\} \setminus P \wedge$
$P \subseteq \{\mathsf{txn}[k'] \mid k' < k \wedge \mathsf{vote}[k'] = \textsc{commit}\} \wedge$
$(\forall t \in T. (\texttt{DECISION}(t, \textsc{commit}) \text{ has been sent})) \wedge$
$(\forall k' < k. \mathsf{vote}[k'] = \textsc{commit} \wedge \mathsf{txn}[k'] \notin T \cup P \implies (\texttt{DECISION}(t, \textsc{abort}) \text{ has been sent})).$

**Proof.** We prove this by induction on the length of the protocol execution. The validity of the above property can be nontrivially affected only by the transitions at lines 19, 26, 44 and 59.

First consider a transition at line 19 that computes $\mathsf{vote}[k]$ as follows:

$\mathsf{vote}[k] = f_s(T, \mathsf{txn}[k]) \sqcap g_s(P, \mathsf{txn}[k]);$
$T = \{\mathsf{txn}[k'] \mid k' < k \wedge \mathsf{phase}[k'] = \textsc{decided} \wedge \mathsf{dec}[k'] = \textsc{commit}\};$
$P = \{\mathsf{txn}[k'] \mid k' < k \wedge \mathsf{phase}[k'] = \textsc{prepared} \wedge \mathsf{vote}[k'] = \textsc{commit}\}.$

Then by Proposition 4,

$T = \{\mathsf{txn}[k'] \mid k' < k \wedge \mathsf{vote}[k'] = \textsc{commit} \wedge d[\mathsf{txn}[k']] = \textsc{commit}\} \setminus P \wedge$
$(\forall t \in T. (\texttt{DECISION}(t, \textsc{commit}) \text{ has been sent})) \wedge$
$(\forall k' < k. \mathsf{vote}[k'] = \textsc{commit} \wedge \mathsf{txn}[k'] \notin T \cup P \implies (\texttt{DECISION}(t, \textsc{abort}) \text{ has been sent})).$

which implies the required.

We next consider the transition at line 44 by a process $p_i$. Then $p_i$ has received messages

$$\texttt{NEW\_LEADER\_ACK}(b'', cballot_j, txn_j, vote_j, dec_j, phase_j)$$

from a quorum of processes $p_j$. Let $b_0 = \max\{cballot_j \mid p_j \in Q'\}$ and $J = \{j \mid cballot_j = b_0\}$.

Consider $k$ such that $\mathsf{vote}[k]$ is defined at $p_i$ after the transition. By induction hypothesis, we have:

$(vote_j[k]$ is defined$) \implies \exists T, P.\, vote_j[k] = f_s(T, txn_j[k]) \sqcap g_s(P, txn_j[k]) \wedge$

$T = \{txn_j[k'] \mid k' < k \wedge vote_j[k'] = \text{COMMIT} \wedge d[t'] = \text{COMMIT}\} \setminus P \wedge$

$P \subseteq \{txn_j[k'] \mid k' < k \wedge vote_j[k'] = \text{COMMIT}\} \wedge$

$(\forall t \in T.\, (\texttt{DECISION}(t, \text{COMMIT})$ has been sent$)) \wedge$

$(\forall k' < k.\, vote_j[k'] = \text{COMMIT} \wedge txn_j[k'] \notin T \cup P \implies (\texttt{DECISION}(t, \text{ABORT})$ has been sent$))$.

From Invariant 2 it follows that

$\forall j_1, j_2 \in J.\, (vote_{j_1}[k']$ is defined$) \wedge (vote_{j_2}[k']$ is defined$) \implies vote_{j_1}[k'] = vote_{j_2}[k']$.

Hence,

$$\exists j_0 \in J.\, \forall k' \leq k.\, \mathsf{vote}[k'] = vote_{j_0}[k'].$$

Then the induction hypothesis implies

$\exists T, P.\, \mathsf{vote}[k] = f_s(T, \mathsf{txn}[k]) \sqcap g_s(P, \mathsf{txn}[k]) \wedge$

$T = \{\mathsf{txn}[k'] \mid k' < k \wedge \mathsf{vote}[k'] = \text{COMMIT} \wedge d[t'] = \text{COMMIT}\} \setminus P \wedge$

$P \subseteq \{\mathsf{txn}[k'] \mid k' < k \wedge \mathsf{vote}[k'] = \text{COMMIT}\} \wedge$

$(\forall t \in T.\, (\texttt{DECISION}(t, \text{COMMIT})$ has been sent$)) \wedge$

$(\forall k' < k.\, \mathsf{vote}[k'] = \text{COMMIT} \wedge \mathsf{txn}[k'] \notin T \cup P \implies (\texttt{DECISION}(t, \text{ABORT})$ has been sent$))$.

as required.

Finally, the cases of the transitions at line 59 and 26 trivially follow from the induction hypothesis. □

We now prove (15)-(17). Take the earliest point in the execution where $d_s[t]$ can be determined as per the definition given earlier. Let $b$ be the ballot used in this definition. Then by Proposition 5 at this point, at the leader of $b$ for some $T_s[t], P_s[t]$ we have

$$\begin{aligned}&d_s[t] = f_s(T_s[t], t) \sqcap g_s(P_s[t], t) \wedge \\ &T_s[t] = \{\mathsf{txn}[k'] \mid k' < k \wedge \mathsf{vote}[k'] = \text{COMMIT} \wedge d[t'] = \text{COMMIT}\} \setminus P_s[t] \wedge \\ &P_s[t] \subseteq \{\mathsf{txn}[k'] \mid k' < k \wedge \mathsf{vote}[k'] = \text{COMMIT}\} \wedge \\ &(\forall t' \in T_s[t].\, (\texttt{DECISION}(t', \text{COMMIT})\ \text{has been sent})) \wedge \\ &(\forall k' < k.\, \mathsf{vote}[k'] = \text{COMMIT} \wedge \mathsf{txn}[k'] \notin T_s[t] \cup P_s[t] \implies \\ &\qquad (\texttt{DECISION}(t', \text{ABORT})\ \text{has been sent})).\end{aligned} \quad (18)$$

But then from the first three conjuncts we get

$$\begin{aligned}&d_s[t] = f_s(T_s[t], t) \sqcap g_s(P_s[t], t) \wedge \\ &T_s[t] = \{t' \mid pos_s[t'] < pos_s[t] \wedge d[t'] = \text{COMMIT}\} \setminus P_s[t] \wedge \\ &P_s[t] \subseteq \{t' \mid pos_s[t'] < pos_s[t] \wedge d_s[t'] = \text{COMMIT}\},\end{aligned}$$

which establishes (15)-(17) for the $T_s[t], P_s[t]$ fixed above.

Finally, let us define a relation $t' \sqsubset_{\text{dec}} t$, signalling that the final decision on transaction $t'$ is taken into account when computing a vote on a transaction $t$:

$t' \sqsubset_{\text{dec}} t \iff \exists s.\, t' \in T_s[t] \vee (pos_s[t'] < pos_s[t] \wedge d_s[t'] = \text{COMMIT} \wedge d[t'] = \text{ABORT} \wedge t' \notin P_s[t])$.

## APPENDIX

Let us also define a relation $t' \sqsubset_{\mathrm{rt}} t$, singalling that transaction $t'$ is decided before transaction $t$ is submitted for certification:

$$t' \sqsubset_{\mathrm{rt}} t \iff \mathtt{decide}(t', \_) \prec_h \mathtt{certify}(t, \_).$$

We now show that the relation $\sqsubset_{\mathrm{dec}} \cup \sqsubset_{\mathrm{rt}}$ is acyclic. To this end, we show that if $t' \sqsubset_{\mathrm{dec}} t$ or $t' \sqsubset_{\mathrm{rt}} t$, then a $\mathtt{DECISION}(t', d[t'])$ message was sent in the execution, and this had happened before any $\mathtt{DECISION}(t, \_)$ message was sent. The case of $t' \sqsubset_{\mathrm{rt}} t$ is obvious and therefore we assume that $t' \sqsubset_{\mathrm{dec}} t$.

Take the earliest point in the execution where we can define $d_s[t]$, and hence, $T_s[t]$ and $P_s[t]$ (by (18)). Then a $\mathtt{DECISION}(t, \_)$ message could not have been sent by this point. If $t' \in T_s[t]$, then by (18) a $\mathtt{DECISION}(t', \mathrm{COMMIT})$ message has been sent earlier. If

$$t' \notin T_s[t] \wedge \mathit{pos}_s[t'] < \mathit{pos}_s[t] \wedge d_s[t'] = \mathrm{COMMIT} \wedge d[t'] = \mathrm{ABORT} \wedge t' \notin P_s[t],$$

then at the above point $\mathsf{txn}[\mathit{pos}_s[t']] = t'$ and $\mathsf{vote}[\mathit{pos}_s[t']] = \mathrm{COMMIT}$, so that by (18) a $\mathtt{DECISION}(t', \mathrm{ABORT})$ message has been sent earlier. We have thus proved that $\sqsubset_{\mathrm{dec}} \cup \sqsubset_{\mathrm{rt}}$ is acyclic.

We now use all the properties established so far to prove the statement of the theorem. For each shard $s$, let us define arrays $\mathit{txn}_s[\,]$ and $\mathit{vote}_s[\,]$ as follows: if $\mathit{pos}_s[t] = k$, then $\mathit{txn}_s[k] = t$ and $\mathit{vote}_s[k] = d_s[t]$. Given (12) and (13), this definition is well-formed. The constraints in Figure 6 and the acyclicity of $\sqsubset_{\mathrm{dec}} \cup \sqsubset_{\mathrm{rt}}$ imply the existence of an execution of the multi-shot 2PC protocol such that it preserves the real-time order of the execution of fault-tolerant protocol and, at its end, the arrays $\mathsf{txn}$ and $\mathsf{vote}$ at a shard $s$ contain $\mathit{txn}_s$ and $\mathit{vote}_s$, respectively. Due to (11), this implies that, if $\mathtt{decide}(t, d)$ occurs in this execution, then $d = d[t]$, as required. □